\documentclass[a4paper,twocolumn]{article}

\usepackage[english]{babel}
\usepackage[utf8]{inputenc}
\usepackage{graphicx}
\usepackage{indentfirst}
\usepackage{float}
\usepackage{fancyhdr}
\usepackage{amssymb}
\usepackage{amsmath}
\usepackage{cite}
\usepackage[labelfont=bf]{caption}
\usepackage{titlesec}

\usepackage[tmargin=1.9cm,lmargin=1.9cm,rmargin=2.54cm,bmargin=2.54cm,
            headheight=23.2pt,headsep=0.4cm]{geometry}
\newcommand{\HRule}{\rule{\linewidth}{0.3mm}}
\usepackage{fancyhdr}
\fancypagestyle{arci}{
  \fancyhf{}
  \fancyhead[C]{\footnotesize{\textit{Sensors \& Transducers Vol 0, Issue 0, Month
                                      2021, pp.}}}
}

\usepackage[dvipsnames]{xcolor}
\usepackage[colorlinks]{hyperref}
\hypersetup{breaklinks=true}
\allowdisplaybreaks

\titleformat*{\section}{\large\bfseries}
\titleformat*{\subsection}{\large\bfseries}
\titlelabel{\thetitle.\ }

\makeatletter
\g@addto@macro\@openbib@code{\setlength{\itemsep}{0pt}}
\makeatother
\begin{document}
\twocolumn[\begin{@twocolumnfalse}
\begin{center}
\Large\textbf{Turbo Coded Single User Massive MIMO}
\end{center}

\begin{center}
{\textbf{$^1$K. Vasudevan, $^1$A. Phani Kumar Reddy, $^1$Gyanesh Kumar Pathak,
         $^2$Mahmoud Albreem}}\\
\small{$^1$Department of Electrical Engineering, Indian Institute of Technology
       Kanpur-208016, India.}\\
\small{$^2$Department of Electrical Engineering, University of Sharjah,
       Sharjah 27272, UAE}\\
\small{$^1$\{vasu, phani, pathak\}@iitk.ac.in}\\
\small{$^2$mahmoud.albreem@asu.edu.om}
\end{center}

\HRule \\
\begin{footnotesize}
\textbf{Summary:} This work deals with turbo coded single user massive multiple
input multiple output (SU-MMIMO) systems, with and without precoding. SU-MMIMO
has a much higher spectral efficiency compared to multi-user massive MIMO (MU-MMIMO)
since independent signals are transmitted from each of the antenna elements (spatial
multiplexing). MU-MMIMO that uses beamforming has a much lower spectral efficiency,
since the same signal (with a delay) is transmitted from each of the antenna elements.
In this work, expressions for the upper bound on the average signal-to-noise ratio
(SNR) per bit and spectral efficiency are derived for SU-MMIMO with and without precoding.
We propose a performance index $f(N_t)$, which is a function of the number of transmit 
antennas $N_t$. Here $f(N_t)$ is the sum of the upper bound on the average SNR per bit
and the spectral efficiency. We demonstrate that when the total number of antennas
($N_{\mathrm{tot}}$) in the transmitter and receiver is fixed, there exists a
minimum value of $f(N_t)$, which has to be avoided. Computer simulations show that the
bit-error-rate (BER) is nearly insensitive to a wide range of the number of transmit
antennas and re-transmissions, when $N_{\mathrm{tot}}$ is large and kept constant.
Thus, the spectral efficiency can be made as large as possible, for a given BER and
$N_{\mathrm{tot}}$.

\vspace{\baselineskip}
\end{footnotesize}
\HRule
\vspace{0.5cm}

\end{@twocolumnfalse}]

\section{Introduction}
\label{Sec:Intro}
Wireless telecommunication standards for 6G and beyond, aim for peak data rates
per user of the order of 100 gigabits per second (Gbps). This can only be achieved
using antenna arrays having a large number of antenna elements at both the
transmitter and receiver (single user massive multiple input multiple output (SU-MMIMO))
and carrier frequencies of the order of terahertz (mmwave frequencies). Much of the
existing literature on massive MIMO deals with multi user case (MU-MMIMO)
\cite{Khwandah2021,9446062,9446676,9444239,9433539,9433520,9432036,9430899,9427230,9417190,9417174,9408648,9408586,9405492,9403922,9400879,9398864,9399122}, where the
base station is equipped with a large number of antennas and each user has only
a single antenna. SU-MMIMO has not yet been studied, excepting
for a few works with equal number of transmit and receive antennas and ideal receiver 
\cite{KV_OpSigPJ2019,73ddc0ea-7d42-4fdd-969d-da08c8e4d0c0},
orthogonal frequency division multiplexing (OFDM)-based practical receiver which 
estimates timing, carrier frequency offset and channel impulse response 
\cite{Vasu_intech:2019,d4bbbdf0-7468-4727-9ebe-76d5e6160b64}, analysis of probability of 
erasure (probability of not detecting an OFDM frame when it is present) 
\cite{KV_SSID2020,9286489e-9169-41d7-9950-1f89bb42fd15}, ideal receiver with unequal
number of transmit and receive antennas with precoding
\cite{KV_ARCI2021,da1844bd-7ee2-4d99-b53b-2339010e03b0}. SU-MMIMO is also presented in
\cite{Vasu_MMIMO_FRTN_2020,Vasu_MMIMO_INGR_2021}.
SU- and MU-MIMO for distributed antenna systems (antennas that are spatially far
apart) is studied in \cite{Schwarz2013}, which is quite different from what is presented
in this work (for example see \textbf{Fig.}~2 of \cite{Vasu_intech:2019}).

Let us look at the differences between SU- and MU-MMIMO 
\cite{Vasu_MMIMO_FRTN_2020,Vasu_MMIMO_INGR_2021}:
\begin{enumerate}
    \item In MU-MMIMO, beamforming is possible only in the downlink, whereas in
          SU-MMIMO, beamforming is possible in both uplink and downlink.
    \item Spatial multiplexing is not possible in MU-MMIMO since the user (mobile
          handset) has only one antenna, whereas in SU-MMIMO spatial multiplexing
          is possible in both uplink and downlink.
\end{enumerate}
The difference between beamforming and spatial multiplexing is enumerated below
\cite{Vasu_MMIMO_FRTN_2020,Vasu_MMIMO_INGR_2021}:
\begin{enumerate}
    \item Beamforming has a lower spectral efficiency, since the same signal
          (with a delay) is transmitted from a large number of antenna elements.
          Spatial multiplexing has a higher spectral efficiency, since independent
          signals are transmitted from a large number of antenna elements.
    \item Beamforming yields a highly directive pencil beam. Spatial multiplexing
          requires a rich scattering channel for effective operation and has no
          directivity.
\end{enumerate}
This work describes two methods of implementing SU-MMIMO with unequal number of
transmit and receive antennas, namely:
\begin{enumerate}
    \item With precoding \cite{KV_ARCI2021,da1844bd-7ee2-4d99-b53b-2339010e03b0}.
    \item Without precoding.
\end{enumerate}
SU-MMIMO without precoding and equal number of transmit and receive antennas has
been described earlier in \cite{KV_OpSigPJ2019,73ddc0ea-7d42-4fdd-969d-da08c8e4d0c0}.
We now briefly discuss the topic of precoding.

Precoding at the transmitter is a technique that dates back to the era of voiceband
modems or wired communications \cite{6768299,49851,93096,93098,120349,119688,237880}.
The term ``precoding'' is quite generic and refers to one or more of the many
different functionalities, as given below:
\begin{enumerate}
    \item It compensates for the distortion introduced by the channel. Note that
          channel compensation at the receiver is referred to as equalization
          \cite{6773303,1457566,1182531,VASUDEVAN20042271,Singer04,Vasu07,Vasu_Book10}.
          Here, channel compensation implies
          removal or minimization of intersymbol interference (ISI).
    \item It performs error control coding, besides channel compensation.
    \item It shapes the spectrum of the transmitted signal, and renders it
          suitable for propagation over the physical channel. Note that most
          channels do not propagate a dc signal and precoding is used to remove
          the dc component in the message signal. At this point, it is important
          to distinguish between a message signal and the transmitted signal.
\end{enumerate}
In the context of wireless multiple input, multiple output (MIMO) systems, the
main task of the precoder is to remove interchannel interference (ICI), either for
single-user or multi-user case
\cite{8169014,9024294,9015969,9025051,9027848,9031293,9040266}. It should
be observed that precoding requires knowledge of the channel state information (CSI)
at the transmitter, which is usually fed back by the receiver to the transmitter. The
receiver estimates CSI from a known training signal that is sent by the transmitter.
CSI usually refers to the channel impulse response (CIR) or its statistics
(mean and covariance), depending on the type of precoder used. Thus,
precoding requires the channel to be time invariant or wide sense
stationary (WSS) over at least one transmit and receive duration. Moreover, precoding
can only be performed on systems employing time division duplex (TDD),
which is a method of half duplex telecommunication. In other words, the channel needs
to be reciprocal, that is, the CIR from the transmitter to receiver must be
identical to that from receiver to transmitter.

In this work, we describe an elegant precoding method which reduces ICI in single
user massive MIMO systems and compare it with the case without precoding
\cite{KV_OpSigPJ2019,Vasu_intech:2019}. Rayleigh flat fading channel is assumed.
If the channel is frequency selective, orthogonal frequency division multiplexing (OFDM)
can be used e.g. single input single output (SISO) OFDM
\cite{Vasu_Book10,6663392,c7888430-cbc2-4c14-87bb-b52780478d85}, single input multiple
output OFDM \cite{Vasudevan2015} or MIMO OFDM
\cite{Vasu_ICWMC2016,Vasu_Adv_Tele_2017,Vasu_intech:2019,d4bbbdf0-7468-4727-9ebe-76d5e6160b64}.

This work is organized as follows. Section~\ref{Sec:Signal_Model} describes the
signal model with and without precoding. In Section~\ref{Sec:Precoding} precoding
for SU-MMIMO is discussed. The case without precoding is presented in 
Section~\ref{Sec:No_Precode}. Section~\ref{Sec:Results} presents the simulation results
and Section~\ref{Sec:Conclude} gives the conclusions.
\section{Signal Model}
\label{Sec:Signal_Model}
\begin{figure*}[tbhp]
\begin{center}
\input{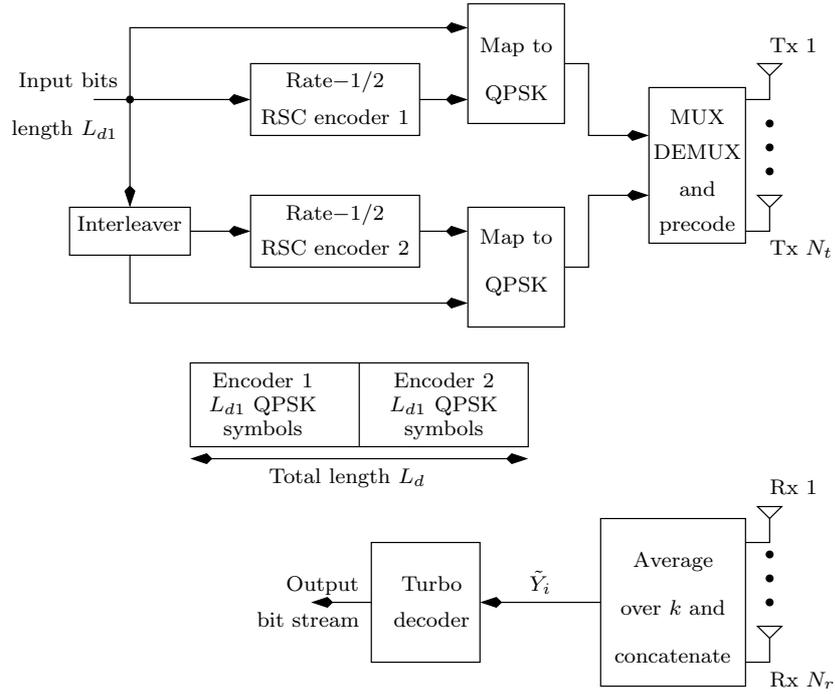}
\caption{System model with precoding.}
\label{Fig:Pap14_System}
\end{center}
\end{figure*}

Consider a precoded MIMO system with $N_t$ transmit and $N_r$ receive antennas, as
shown in \textbf{Fig.}~\ref{Fig:Pap14_System} \cite{KV_ARCI2021}.
The precoded received signal in the $k^{th}$ ($0\leq k\leq N_{rt}-1$,
$k$ is an integer), re-transmission is given by
\begin{equation}
\label{Eq:Pre_Massive_MIMO_Eq1}
\tilde{\mathbf{R}}_k = \tilde{\mathbf{H}}_k
                       \tilde{\mathbf{H}}_k^H
                       \mathbf{S}^p +
                       \tilde{\mathbf{W}}_k
\end{equation}
where $\tilde{\mathbf{R}}_k\in \mathbb{C}^{N_r\times 1}$ is the received
vector, $\tilde{\mathbf{H}}_k\in \mathbb{C}^{N_r\times N_t}$ is the channel
matrix and $\tilde{\mathbf{W}}_k\in \mathbb{C}^{N_r\times 1}$ is the additive 
white Gaussian noise (AWGN) vector. The transmitted symbol vector is
$\mathbf{S}^p\in \mathbb{C}^{N_r\times 1}$, whose elements are drawn from an
$M$-ary constellation. Boldface letters denote vectors or matrices. Complex
quantities are denoted by a  tilde. However tilde is not used for complex
symbols $\mathbf{S}^p$. The elements of $\tilde{\mathbf{H}}_{k}$ are
statistically independent, zero mean, circularly symmetric complex Gaussian with
variance per dimension equal to
$\sigma_{H}^{2}$, as given by (2) of \cite{KV_OpSigPJ2019}. Similarly, the elements of
$\tilde{\mathbf{W}}_k$ are statistically independent, zero mean, circularly
symmetric complex Gaussian with 
variance per dimension equal to $\sigma_W^2$, as given by (3) of \cite{KV_OpSigPJ2019}.

\begin{figure*}[tbhp]
\begin{center}
\input{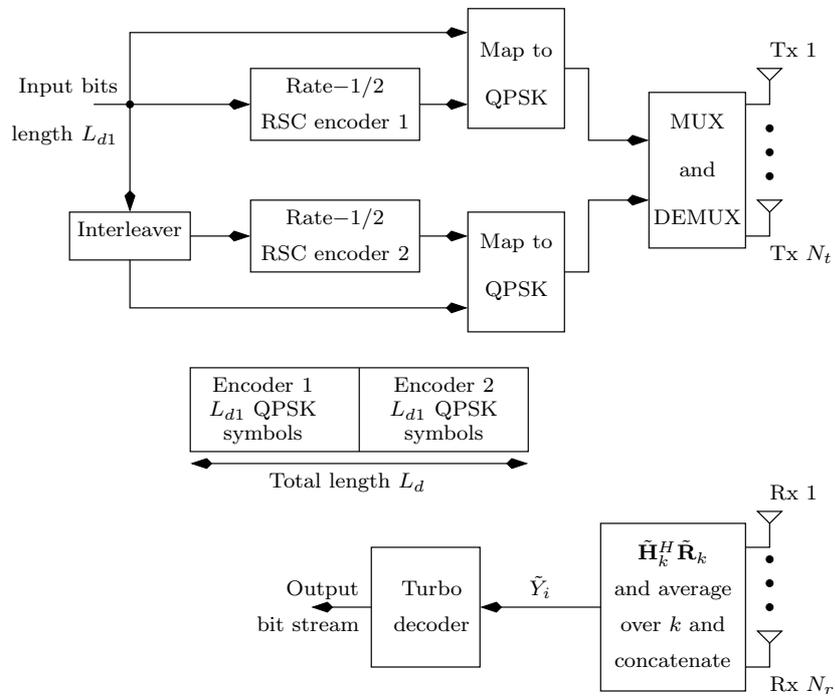}
\caption{System model without precoding.}
\label{Fig:No_Precode_System}
\end{center}
\end{figure*}
The system model without precoding is shown in
Figure~\ref{Fig:No_Precode_System} which is similar to Figure~1 of
\cite{KV_OpSigPJ2019}, excepting that here the number of transmit antennas is not
equal to the number of receive antennas.
The received signal without precoding, in the $k^{th}$ re-transmission is given
by (see also (1) of \cite{KV_OpSigPJ2019})
\begin{equation}
\label{Eq:Pre_Massive_MIMO_Eq2}
\tilde{\mathbf{R}}_k = \tilde{\mathbf{H}}_k
                       \mathbf{S} +
                       \tilde{\mathbf{W}}_k
\end{equation}
where $\mathbf{S}\in \mathbb{C}^{N_t\times 1}$ whose elements are drawn
from an $M$-ary constellation. In this work, the elements of $\mathbf{S}^p$ and
$\mathbf{S}$ are turbo coded and mapped to a QPSK constellation with
coordinates $\pm 1 \pm\mathrm{j}$, as depicted
in \textbf{Fig.}~\ref{Fig:Pap14_System}.
Moreover, here $\tilde{\mathbf{H}}_k$ is an
$N_r\times N_t$ matrix, whereas in \cite{KV_OpSigPJ2019} $\tilde{\mathbf{H}}_k$ is
an $N\times N$ matrix. We assume that $\tilde{\mathbf{H}}_k$ and
$\tilde{\mathbf{W}}_k$ are independent across re-transmissions, hence (4) in 
\cite{KV_OpSigPJ2019} is valid with $N$ replaced by $N_r$.
We now proceed to analyze the signal models in
(\ref{Eq:Pre_Massive_MIMO_Eq1}) and (\ref{Eq:Pre_Massive_MIMO_Eq2}).
\section{Precoding}
\label{Sec:Precoding}
The $i^{th}$ element of $\tilde{\mathbf{R}}_k$ in (\ref{Eq:Pre_Massive_MIMO_Eq1}) is
\begin{equation}
\label{Eq:Pre_Massive_MIMO_Eq3}
\tilde{R}_{k,\, i} = \tilde{F}_{k,\, i,\, i} S_i +
                     \tilde{I}_{k,\, i} +
                     \tilde{W}_{k,\, i}
                     \quad \mbox{for $1\leq i\leq N_r$}
\end{equation}
where
\begin{align}
\label{Eq:Pre_Massive_MIMO_Eq4}
\tilde{F}_{k,\, i,\, i} & = \sum_{j=1}^{N_t}
                            \left|
                            \tilde{H}_{k,\, i,\, j}
                            \right|^2                         \nonumber  \\ 
\tilde{I}_{k,\, i}      & = \sum_{\substack{j=1\\j\neq i}}^{N_r}
                            \tilde{F}_{k,\, i,\, j} S_j       \nonumber  \\
\tilde{F}_{k,\, i,\, j} & = \sum_{l=1}^{N_t}
                            \tilde{H}_{k,\, i,\, l}
                            \tilde{H}_{k,\, j,\, l}^* \qquad \mbox{for $i\ne j$}.
\end{align}
The desired signal in (\ref{Eq:Pre_Massive_MIMO_Eq3}) is $F_{k,\, i,\, i}S_i$, the
interference term is $\tilde{I}_{k,\, i}$ and the noise term is
$\tilde{W}_{k,\, i}$. Now
\begin{align}
\label{Eq:Pre_Massive_MIMO_Eq5}
 E
\left[
\tilde{F}_{k,\, i,\, i}^2
\right] & =  E
            \left[
            \sum_{j=1}^{N_t}
            \left|
            \tilde{H}_{k,\, i,\, j}
            \right|^2
            \sum_{l=1}^{N_t}
            \left|
            \tilde{H}_{k,\, i,\, l}
            \right|^2
            \right]                              \nonumber  \\
        & =  E
            \left[
            \sum_{j=1}^{N_t}
            \tilde{H}_{k,\, i,\, j,\, I}^2 +
            \tilde{H}_{k,\, i,\, j,\, Q}^2
            \right.                              \nonumber  \\
        &   \qquad
            \times \left.
            \sum_{l=1}^{N_t}
            \tilde{H}_{k,\, i,\, l,\, I}^2 +
            \tilde{H}_{k,\, i,\, l,\, Q}^2
            \right]                              \nonumber  \\
        & = 4\sigma_H^4 N_t (N_t+1)
\end{align}
where the subscript ``$I$'' denotes the in-phase part and the subscript ``$Q$''
denotes the quadrature part of a complex quantity and the following relation
has been used \cite{Papoulis91,Vasu_AC_PS}
\begin{equation}
\label{Eq:Pre_Massive_MIMO_Eq6}
 E
\left[
 X^4
\right] = 3 \sigma_X^4
\end{equation}
where $X$ is a zero-mean, real-valued Gaussian random variable with variance
$\sigma^2_X$. Moreover from (\ref{Eq:Pre_Massive_MIMO_Eq4}) and (2) in
\cite{KV_OpSigPJ2019}
\begin{equation}
\label{Eq:Pre_Massive_MIMO_Eq6_1}
 E
\left[
\tilde{F}_{k,\, i,\, i}
\right] = 2 \sigma_H^2 N_t.
\end{equation}
\begin{figure*}[tbhp]
\begin{center}
\input{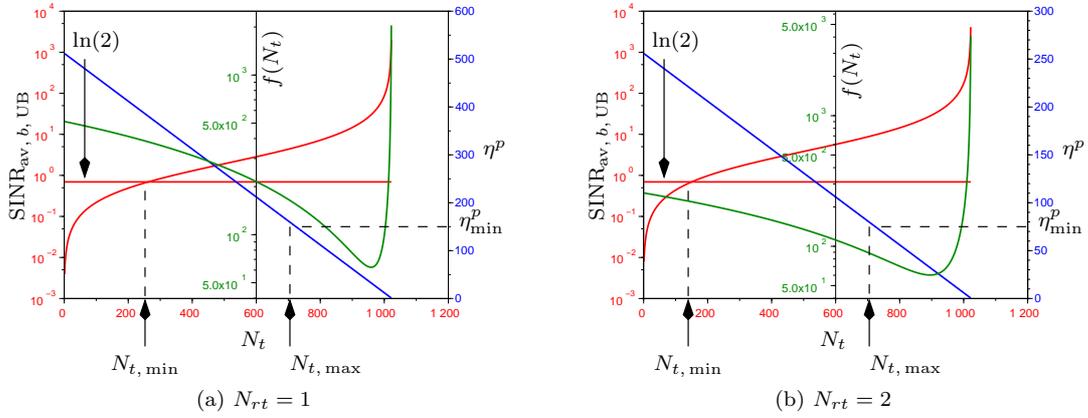}
\caption{$\mathrm{SINR}_{\mathrm{av},\, b,\,\mathrm{UB}}$ and $\eta^p$ as a
         function of $N_t$ for $N_{\mathrm{tot}}=1024$, with precoding.}
\label{Fig:Pre_SINR_UB_Eta_p}
\end{center}
\end{figure*}
We also have
\begin{align}
\label{Eq:Pre_Massive_MIMO_Eq7}
 E
\left[
\left|
 \tilde{I}_{k,\, i}
\right|^2
\right] & =  E
            \left[
            \sum_{\substack{j=1\\j\neq i}}^{N_r}
            \tilde{F}_{k,\, i,\, j} S_j
            \right.                                     \nonumber  \\
        &   \qquad \times
            \left.
            \sum_{\substack{l=1\\l\neq i}}^{N_r}
            \tilde{F}_{k,\, i,\, l}^* S_l^*
            \right]                                     \nonumber  \\
        & = \sum_{\substack{j=1\\j\ne i}}^{N_r}
            \sum_{\substack{l=1\\l\ne i}}^{N_r}
             P_{\mathrm{av}}
             E
            \left[
            \tilde{F}_{k,\, i,\, j}
            \tilde{F}_{k,\, i,\, l}^*
            \right]
            \delta_K(j-l)                               \nonumber  \\
        & =  P_{\mathrm{av}}
            \sum_{\substack{j=1\\j\ne i}}^{N_r}
             E
            \left[
            \left|
            \tilde{F}_{k,\, i,\, j}
            \right|^2
            \right]
\end{align}
where $\delta_K(\cdot)$ is the Kronecker delta function 
\cite{Vasu_Book10,KV_OpSigPJ2019}, we have assumed
independence between $\tilde{F}_{k,\, i,\, j}$ and $S_j$ and \cite{KV_OpSigPJ2019}
\begin{align}
\label{Eq:Pre_Massive_MIMO_Eq8}
 E
\left[
 S_j S_l^*
\right] & = P_{\mathrm{av}} \delta_K(j-l)    \nonumber  \\
        & = 2 \delta_K(j-l).
\end{align}
Now
\begin{align}
\label{Eq:Pre_Massive_MIMO_Eq9}
 E
\left[
\left|
\tilde{F}_{k,\, i,\, j}
\right|^2
\right] & =  E
            \left[
            \sum_{l=1}^{N_t}
            \tilde{H}_{k,\, i,\, l}
            \tilde{H}_{k,\, j,\, l}^* 
            \right.                              \nonumber  \\
        &   \qquad \times
            \left.
            \sum_{m=1}^{N_t}
            \tilde{H}_{k,\, i,\, m}^*
            \tilde{H}_{k,\, j,\, m}
            \right]                              \nonumber  \\
        & = \sum_{l=1}^{N_t}
            \sum_{m=1}^{N_t}
             E
            \left[
            \tilde{H}_{k,\, i,\, l}
            \tilde{H}_{k,\, i,\, m}^*
            \right]                              \nonumber  \\
        &   \qquad \times
             E
            \left[
            \tilde{H}_{k,\, j,\, m}
            \tilde{H}_{k,\, j,\, l}^*
            \right]                              \nonumber  \\
        & = \sum_{l=1}^{N_t}
            \sum_{m=1}^{N_t}
             4
            \sigma_H^4
            \delta_K(l-m)                        \nonumber  \\
        & =  4
            \sigma_H^4 N_t.
\end{align}
Substituting (\ref{Eq:Pre_Massive_MIMO_Eq9}) in (\ref{Eq:Pre_Massive_MIMO_Eq7})
and using (\ref{Eq:Pre_Massive_MIMO_Eq8}) we get
\begin{equation}
\label{Eq:Pre_Massive_MIMO_Eq10}
 E
\left[
\left|
\tilde{I}_{k,\, i}
\right|^2
\right] = 8 \sigma_H^4 N_t (N_r-1).
\end{equation}
Due to independence between $\tilde{I}_{k,\, i}$ and $\tilde{W}_{k,\, i}$ in
(\ref{Eq:Pre_Massive_MIMO_Eq3}) we have from (\ref{Eq:Pre_Massive_MIMO_Eq10})
and (3) of \cite{KV_OpSigPJ2019}
\begin{align}
\label{Eq:Pre_Massive_MIMO_Eq11}
 E
\left[
\left|
\tilde{I}_{k,\, i} + \tilde{W}_{k,\, i}
\right|^2
\right] & =  E
            \left[
            \left|
            \tilde{I}_{k,\, i}
            \right|^2
            \right] +
             E
            \left[
            \left|
            \tilde{W}_{k,\, i}
            \right|^2
            \right]                                    \nonumber  \\
        & =  8
            \sigma_H^4 N_t (N_r-1) + 2\sigma_W^2     \nonumber  \\
        & = \sigma^2_{U'}                              \qquad \mbox{(say)}.
\end{align}
Now, each element of $\mathbf{S}^p$ in (\ref{Eq:Pre_Massive_MIMO_Eq1}) carries
$1/(2N_{rt})$ bits of information \cite{KV_OpSigPJ2019}. Therefore, each element
of $\tilde{\mathbf{R}}_k$ also carries $1/(2N_{rt})$ bits of information. Hence,
the average signal to interference plus noise ratio per bit of $\tilde{R}_{k,\, i}$ in
(\ref{Eq:Pre_Massive_MIMO_Eq3}) is defined as, using (\ref{Eq:Pre_Massive_MIMO_Eq5}),
(\ref{Eq:Pre_Massive_MIMO_Eq8}) and (\ref{Eq:Pre_Massive_MIMO_Eq11})
\begin{align}
\label{Eq:Pre_Massive_MIMO_Eq12}
\mathrm{SINR}_{\mathrm{av},\, b}
        & = \frac{
             E
            \left[
            \left|
            \tilde{F}_{k,\, i,\, i} S_i
            \right|^2
            \right]
            \times 2N_{rt}}
            {
             E
           \left[
           \left|
           \tilde{I}_{k,\, i} + \tilde{W}_{k,\, i}
           \right|^2
           \right]
           }                                               \nonumber  \\
        & = \frac{8\sigma_H^4 N_t (N_t+1)\times 2 N_{rt}}
            {
             8\sigma_H^4 N_t (N_r-1) + 2\sigma_W^2
            }.
\end{align}
\begin{figure*}[tbhp]
\begin{center}
\input{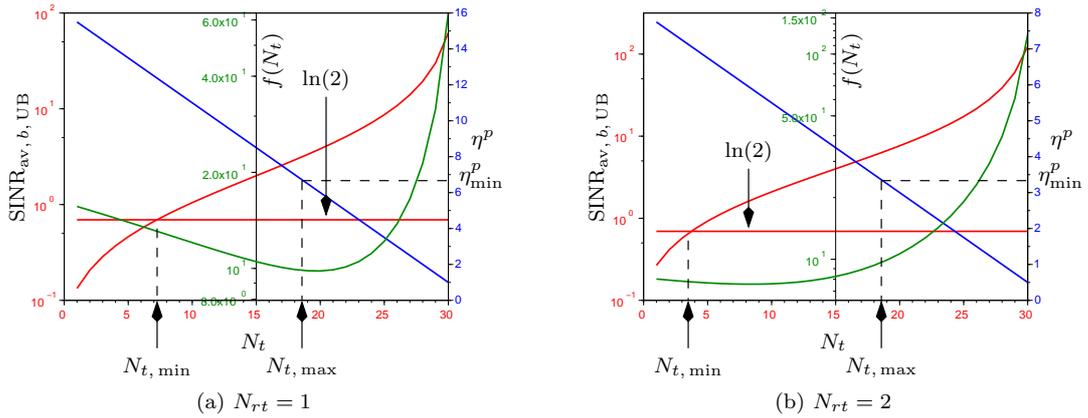}
\caption{$\mathrm{SINR}_{\mathrm{av},\, b,\,\mathrm{UB}}$ and $\eta^p$ as a
         function of $N_t$ for $N_{\mathrm{tot}}=32$, with precoding.}
\label{Fig:Pre_SINR_UB_Eta_p_Ntot32}
\end{center}
\end{figure*}
When $\sigma_W^2=0$ in (\ref{Eq:Pre_Massive_MIMO_Eq12}), we get the upper bound
on $\mathrm{SINR}_{\mathrm{av},\, b}$ as given below
\begin{align}
\label{Eq:Pre_Massive_MIMO_Eq13}
\mathrm{SINR}_{\mathrm{av},\, b,\,\mathrm{UB}}
& = \frac{8\sigma_H^4 N_t (N_t+1)\times 2 N_{rt}}
         {8\sigma_H^4 N_t (N_r-1)}                  \nonumber  \\
& = \frac{2N_{rt}(N_t+1)}{N_r-1}.
\end{align}
The information contained in $\mathbf{S}^p$ in (\ref{Eq:Pre_Massive_MIMO_Eq1})
is $N_r/(2N_{rt})$ bits. Hence the spectral efficiency of the precoded system is
\begin{equation}
\label{Eq:Pre_Massive_MIMO_Eq14}
\eta^p = \frac{N_r}{2N_{rt}} \qquad \mbox{bits per transmission}.
\end{equation}
Note that both (\ref{Eq:Pre_Massive_MIMO_Eq13}) and (\ref{Eq:Pre_Massive_MIMO_Eq14})
need to be as large as possible to minimize the BER and maximize the spectral
efficiency. Let
\begin{equation}
\label{Eq:Pre_Massive_MIMO_Eq15}
N_{\mathrm{tot}} = N_t+N_r.
\end{equation}
Define
\begin{align}
\label{Eq:Pre_Massive_MIMO_Eq16}
f(N_t) & = \mathrm{SINR}_{\mathrm{av},\, b,\,\mathrm{UB}} + \eta^p  \nonumber  \\
       & = \frac{2N_{rt}(N_t+1)}{N_r-1} + \frac{N_r}{2N_{rt}}       \nonumber  \\
       & = \frac{2N_{rt}(N_t+1)}{N_{\mathrm{tot}}-N_t-1} +
           \frac{N_{\mathrm{tot}}-N_t}{2N_{rt}}
\end{align}
where we have used (\ref{Eq:Pre_Massive_MIMO_Eq15}).
We need to find $N_t$ such that $f(N_t)$ is maximized. The plot of
$\mathrm{SINR}_{\mathrm{av},\, b,\,\mathrm{UB}}$ (red curve), $\eta^p$ (blue curve)
and $f(N_t)$ (green curve), as a function of $N_t$, keeping $N_{\mathrm{tot}}$
fixed, is depicted in \textbf{Fig.}~\ref{Fig:Pre_SINR_UB_Eta_p} and
\ref{Fig:Pre_SINR_UB_Eta_p_Ntot32}. Note that
$\mathrm{SINR}_{\mathrm{av},\, b,\,\mathrm{UB}}$ increases monotonically and $\eta^p$
decreases monotonically, with increasing $N_t$. We also find that
$f(N_t)$ has a minimum (not maximum) at
\begin{equation}
\label{Eq:Pre_Massive_MIMO_Eq17}
N_t = N_{\mathrm{tot}}-2N_{rt}\sqrt{N_{\mathrm{tot}}} -1
\end{equation}
which is obtained by differentiating $f(N_t)$ in (\ref{Eq:Pre_Massive_MIMO_Eq16})
with respect to $N_t$ and setting the result to zero. Therefore, the only possible
solution is to avoid the minimum.  Clearly we require
$\mathrm{SINR}_{\mathrm{av},\, b,\,\mathrm{UB}}>\ln(2)$, since it is the minimum
average SNR per bit required for error-free transmission over any type of channel
\cite{KV_OpSigPJ2019}. We also require $\eta^p > \eta^p_{\mathrm{min}}$, where
$\eta^p_{\mathrm{min}}$ is chosen by the system designer. Thus, we arrive at a
range of the number of transmit antennas
($N_{t,\,\mathrm{min}} \le N_t \le N_{t,\,\mathrm{max}}$) that can be used, as shown
in \textbf{Fig.}~\ref{Fig:Pre_SINR_UB_Eta_p} and
\ref{Fig:Pre_SINR_UB_Eta_p_Ntot32}. Note that in
\textbf{Fig.}~\ref{Fig:Pre_SINR_UB_Eta_p_Ntot32}(b) the minimum of
$f(N_t)$ cannot be avoided, since $\eta^p_{\mathrm{min}}$ would be too small.

Next, similar to (20) in \cite{KV_OpSigPJ2019}, consider
\begin{align}
\label{Eq:Pre_Massive_MIMO_Eq18}
\tilde{Y}_i & = \frac{1}{N_{rt}}
                \sum_{k=0}^{N_{rt}-1}
                \tilde{R}_{k,\, i}                      \nonumber  \\
            & = \frac{1}{N_{rt}}
                \sum_{k=0}^{N_{rt}-1}
                \left(
                \tilde{F}_{k,\, i,\, i} S_i +
                \tilde{I}_{k,\, i} +
                \tilde W_{k,\, i}
                \right)                                \nonumber  \\
            & =  F_i S_i + \tilde{U}_i \qquad \mbox{for $1\le i \le N_r$}
\end{align}
where $\tilde{R}_{k,\, i}$ is given by (\ref{Eq:Pre_Massive_MIMO_Eq3}), $F_i$
is real-valued and
\begin{align}
\label{Eq:Pre_Massive_MIMO_Eq19}
F_i         & = \frac{1}{N_{rt}}
                \sum_{k=0}^{N_{rt}-1}
                \tilde{F}_{k,\, i,\, i}                        \nonumber  \\
\tilde{U}_i & = \frac{1}{N_{rt}}
                \sum_{k=0}^{N_{rt}-1}
                \left(
                \tilde{I}_{k,\, i} +
                \tilde{W}_{k,\, i}
                \right)                                        \nonumber  \\
            & = \frac{1}{N_{rt}}
                \sum_{k=0}^{N_{rt}-1}
                \tilde{U}_{k,\, i}'   \qquad \mbox{(say)}.
\end{align}
\begin{figure*}[tbhp]
\begin{center}
\input{pap16_ntot32.pstex_t}
\caption{$\mathrm{SINR}_{\mathrm{av},\, b,\,\mathrm{UB}}$ and $\eta$ as a
         function of $N_t$ for $N_{\mathrm{tot}}=32$, without precoding.}
\label{Fig:Pap16_Ntot32}
\end{center}
\end{figure*}
Since $\tilde{F}_{k,\, i,\, i}$ and $\tilde{U}_{k,\, i}'$ are statistically
independent over re-transmissions ($k$), we have
\begin{align}
\label{Eq:Pre_Massive_MIMO_Eq20}
 E
\left[
F_i^2
\right]     & = \frac{1}{N_{rt}^2}
                 E
                \left[
                \sum_{k=0}^{N_{rt}-1}
                \tilde{F}_{k,\, i,\, i}
                \sum_{n=0}^{N_{rt}-1}
                \tilde{F}_{n,\, i,\, i}
                \right]                                        \nonumber  \\
            & = \frac{4\sigma_H^4 N_t
                \left[N_t+1+N_t(N_{rt}-1)
                \right]}{N_{rt}}                               \nonumber  \\
            & = \frac{4\sigma_H^4 N_t(N_t N_{rt}+1)}
                     {N_{rt}}                                  \nonumber  \\
 E
\left[
\left|
\tilde{U}_i
\right|^2
\right]     & = \frac{\sigma^2_{U'}}{N_{rt}}                   \nonumber  \\
            & = \frac{8 \sigma_H^4 N_t (N_r-1) + 2\sigma_W^2}{N_{rt}}
\end{align}
where we have used (\ref{Eq:Pre_Massive_MIMO_Eq5}),
(\ref{Eq:Pre_Massive_MIMO_Eq6_1}),
(\ref{Eq:Pre_Massive_MIMO_Eq11}) and the fact that
\begin{equation}
\label{Eq:Pre_Massive_MIMO_Eq21}
 E
\left[
\tilde{U}_{k,\, i}'
\right] = 0
\end{equation}
where $\tilde{U}_{k,\, i}'$ is defined in (\ref{Eq:Pre_Massive_MIMO_Eq19}). Next,
we compute the average SINR per bit for $\tilde{Y}_i$ in
(\ref{Eq:Pre_Massive_MIMO_Eq18}). Note that since $\tilde{Y}_i$ is a ``combination''
of $N_{rt}$ re-transmissions, its information content is $N_{rt}/(2N_{rt})=1/2$ bit
(recall that the information content of $\tilde{R}_{k,\, i}$ in
(\ref{Eq:Pre_Massive_MIMO_Eq18}) is $1/(2N_{rt})$ bits). Therefore
\begin{align}
\label{Eq:Pre_Massive_MIMO_Eq22}
\mathrm{SINR}_{\mathrm{av},\, b,\, C}
& = \frac{E\left[\left|F_i S_i\right|^2\right]\times 2}
         {E\left[\left|\tilde{U}_i\right|^2\right]}             \nonumber  \\
& = \frac{8\sigma_H^4 N_t (N_t N_{rt}+1)\times 2}
         {8\sigma_H^4 N_t (N_r-1)+2\sigma^2_W}
\end{align}
where the subscript ``$C$'' denotes ``after combining'' and we have used
(\ref{Eq:Pre_Massive_MIMO_Eq8}) and (\ref{Eq:Pre_Massive_MIMO_Eq20}). Note that
we prefer to use the word ``combining'' rather than averaging, since it is
more appropriate in terms of the ``information content'' in $\tilde{Y}_i$. Once
again with $\sigma^2_W=0$ and $N_t N_{rt}\gg 1$ we get the approximate upper
bound on $\mathrm{SINR}_{\mathrm{av},\, b,\, C}$ as
\begin{align}
\label{Eq:Pre_Massive_MIMO_Eq23}
\mathrm{SINR}_{\mathrm{av},\, b,\, C,\,\mathrm{UB}}
& = \frac{8\sigma_H^4 N_t (N_t N_{rt}+1)\times 2}
         {8\sigma_H^4 N_t (N_r-1)}                         \nonumber  \\
& \approx
    \frac{2N_{rt}N_t}{N_r-1}                               \nonumber  \\
& \approx
    \mathrm{SINR}_{\mathrm{av},\, b,\,\mathrm{UB}}
\end{align}
when $N_t\gg 1$. Thus, the upper bound on the average SINR per bit before and
after combining are nearly identical. Observe that re-transmitting the data increases
the upper bound on the average SINR per bit, it does not improve the BER performance,
which is seen in Section~\ref{Sec:Results}. After concatenation, the signal
$\tilde Y_i$
in (\ref{Eq:Pre_Massive_MIMO_Eq18}) for $0\le i\le L_d-1$ is sent to the turbo
decoder. The details of turbo decoding will not be discussed here.
\begin{figure*}[tbhp]
\begin{center}
\input{pap16_ntot1024.pstex_t}
\caption{$\mathrm{SINR}_{\mathrm{av},\, b,\,\mathrm{UB}}$ and $\eta$ as a
         function of $N_t$ for $N_{\mathrm{tot}}=1024$, without precoding.}
\label{Fig:Pap16_Ntot1024}
\end{center}
\end{figure*}
\section{No Precoding}
\label{Sec:No_Precode}
The block diagram of the system without precoding in similar to
\textbf{Fig.}~1 in \cite{KV_OpSigPJ2019}, excepting that now there are $N_t$
transmit and $N_r$ receive antennas.
The $i^{th}$ element of $\tilde{\mathbf{H}}_k^H\tilde{\mathbf{R}}_k$, where
$\tilde{\mathbf{R}_k}$ is given by (\ref{Eq:Pre_Massive_MIMO_Eq2}), is
(similar to (10) of \cite{KV_OpSigPJ2019})
\begin{equation}
\label{Eq:Pre_Massive_MIMO_Eq23_0}
\tilde{Y}_{k,\, i} = \tilde{F}_{k,\, i,\, i} S_i +
                     \tilde{I}_{k,\, i} +
                     \tilde{V}_{k,\, i}
                     \quad \mbox{for $1\leq i\leq N_t$}
\end{equation}
where
\begin{align}
\label{Eq:Pre_Massive_MIMO_Eq23_1}
\tilde{V}_{k,\, i}      & = \sum_{j=1}^{N_r}
                            \tilde{H}_{k,\, j,\, i}^*
                            \tilde W_{k,\, j}                 \nonumber  \\ 
\tilde{I}_{k,\, i}      & = \sum_{\substack{j=1\\j\neq i}}^{N_t}
                            \tilde{F}_{k,\, i,\, j} S_j       \nonumber  \\
\tilde{F}_{k,\, i,\, j} & = \sum_{l=1}^{N_r}
                            \tilde{H}_{k,\, l,\, i}^*
                            \tilde{H}_{k,\, l,\, j}.
\end{align}
It can be shown that
\begin{equation}
\label{Eq:Pre_Massive_MIMO_Eq23_2}
 E
\left[
\tilde{F}_{k,\, i,\, i}^2
\right] = 4\sigma_H^4 N_r (N_r+1).
\end{equation}
We also have
\begin{equation}
\label{Eq:Pre_Massive_MIMO_Eq23_3}
 E
\left[
\tilde{F}_{k,\, i,\, i}
\right] = 2 \sigma_H^2 N_r
\end{equation}
and
\begin{align}
\label{Eq:Pre_Massive_MIMO_Eq23_4}
 E
\left[
\left|
\tilde{I}_{k,\, i}
\right|^2
\right] & = 8\sigma_H^4 N_r (N_t-1)                   \nonumber  \\
 E
\left[
\left|
\tilde{V}_{k,\, i}
\right|^2
\right] & = 4\sigma^2_W\sigma_H^2 N_r.
\end{align}
The total power of interference plus noise is
\begin{align}
\label{Eq:Pre_Massive_MIMO_Eq23_5}
 E
\left[
\left|
\tilde{I}_{k,\, i} + \tilde{V}_{k,\, i}
\right|^2
\right] & =  E
            \left[
            \left|
            \tilde{I}_{k,\, i}
            \right|^2
            \right] +
             E
            \left[
            \left|
            \tilde{V}_{k,\, i}
            \right|^2
            \right]                                    \nonumber  \\
        & =  8
            \sigma_H^4 N_r (N_t-1)                     \nonumber  \\
        &   \qquad +
             4
            \sigma_W^2\sigma^2_H N_r                   \nonumber  \\
        & = \sigma^2_{U'}                              \qquad \mbox{(say)}.
\end{align}
Observe that the total information emitted by $N_t$ antennas per transmission
is $N_t/(2N_{rt})$ bits. Therefore, the information contained in
$\tilde Y_{k,\, i}$ in (\ref{Eq:Pre_Massive_MIMO_Eq23_0}) is
$N_t/(2N_{rt} N_t)$ bits. Hence, the average SINR per bit is
\begin{align}
\label{Eq:Pre_Massive_MIMO_Eq23_6}
\mathrm{SINR}_{\mathrm{av},\, b}
        & = \frac{
             E
            \left[
            \left|
            \tilde{F}_{k,\, i,\, i} S_i
            \right|^2
            \right]
            \times 2N_{rt}}
            {
             E
           \left[
           \left|
           \tilde{I}_{k,\, i} + \tilde{W}_{k,\, i}
           \right|^2
           \right]
           }                                               \nonumber  \\
        & = \frac{8\sigma_H^4 (N_r+1)\times 2 N_{rt}}
            {
             8\sigma_H^4 (N_t-1) + 4\sigma_W^2 \sigma^2_H
            }
\end{align}
and for $\sigma^2_W=0$ we get the upper bound on SINR per bit as
\begin{equation}
\label{Eq:Pre_Massive_MIMO_Eq23_7}
\mathrm{SINR}_{\mathrm{av},\, b,\,\mathrm{UB}}
 = \frac{(N_r+1)\times 2 N_{rt}}
        {N_t-1}.
\end{equation}
The spectral efficiency without precoding is
\begin{equation}
\label{Eq:Pre_Massive_MIMO_Eq23_8}
\eta = \frac{N_t}{2N_{rt}} \qquad \mbox{bits per transmission}.
\end{equation}
Define
\begin{align}
\label{Eq:Pre_Massive_MIMO_Eq23_9}
f(N_t) & = \mathrm{SINR}_{\mathrm{av},\, b,\,\mathrm{UB}} + \eta  \nonumber  \\
       & = \frac{2N_{rt}(N_r+1)}{N_t-1} + \frac{N_t}{2N_{rt}}     \nonumber  \\
       & = \frac{2N_{rt}(N_{\mathrm{tot}}-N_t+1)}{N_t-1} +
           \frac{N_t}{2N_{rt}}
\end{align}
where $N_{\mathrm{tot}}$ is given by (\ref{Eq:Pre_Massive_MIMO_Eq15}). The
value of $N_t$ that minimizes $f(N_t)$ in (\ref{Eq:Pre_Massive_MIMO_Eq23_9})
is given by
\begin{equation}
\label{Eq:Pre_Massive_MIMO_Eq23_10}
N_t = 2N_{rt}\sqrt{N_{\mathrm{tot}}} +1.
\end{equation}
Thus, we can arrive at a range of transmit antennas that can be used, avoiding
the minimum of $f(N_t)$. This is illustrated in
\textbf{Figs.}~\ref{Fig:Pap16_Ntot32} and \ref{Fig:Pap16_Ntot1024}.

Again similar to (20) in \cite{KV_OpSigPJ2019}, we compute the average of
$\tilde{Y}_{k,\, i}$ over all re-transmissions, as given by
\begin{align}
\label{Eq:Pre_Massive_MIMO_Eq23_11}
\tilde{Y}_i & = \frac{1}{N_{rt}}
                \sum_{k=0}^{N_{rt}-1}
                \tilde{Y}_{k,\, i}                      \nonumber  \\
            & = \frac{1}{N_{rt}}
                \sum_{k=0}^{N_{rt}-1}
                \left(
                \tilde{F}_{k,\, i,\, i} S_i +
                \tilde{I}_{k,\, i} +
                \tilde V_{k,\, i}
                \right)                                \nonumber  \\
            & =  F_i S_i + \tilde{U}_i \qquad \mbox{for $1\le i \le N_t$}
\end{align}
where $\tilde{Y}_{k,\, i}$ is given by (\ref{Eq:Pre_Massive_MIMO_Eq23_0}) and
\begin{align}
\label{Eq:Pre_Massive_MIMO_Eq23_12}
F_i         & = \frac{1}{N_{rt}}
                \sum_{k=0}^{N_{rt}-1}
                \tilde{F}_{k,\, i,\, i}                        \nonumber  \\
\tilde{U}_i & = \frac{1}{N_{rt}}
                \sum_{k=0}^{N_{rt}-1}
                \left(
                \tilde{I}_{k,\, i} +
                \tilde{V}_{k,\, i}
                \right)                                        \nonumber  \\
            & = \frac{1}{N_{rt}}
                \sum_{k=0}^{N_{rt}-1}
                \tilde{U}_{k,\, i}'   \qquad \mbox{(say)}.
\end{align}
Since $\tilde{F}_{k,\, i,\, i}$ and $\tilde{U}_{k,\, i}'$ are statistically
independent over re-transmissions ($k$), we have
\begin{align}
\label{Eq:Pre_Massive_MIMO_Eq23_13}
 E
\left[
F_i^2
\right]     & = \frac{1}{N_{rt}^2}
                 E
                \left[
                \sum_{k=0}^{N_{rt}-1}
                \tilde{F}_{k,\, i,\, i}
                \sum_{n=0}^{N_{rt}-1}
                \tilde{F}_{n,\, i,\, i}
                \right]                                        \nonumber  \\
            & = \frac{4\sigma_H^4 N_r
                \left[N_r+1+N_r(N_{rt}-1)
                \right]}{N_{rt}}                               \nonumber  \\
            & = \frac{4\sigma_H^4 N_r(N_r N_{rt}+1)}
                     {N_{rt}}                                  \nonumber  \\
 E
\left[
\left|
\tilde{U}_i
\right|^2
\right]     & = \frac{\sigma^2_{U'}}{N_{rt}}                   \nonumber  \\
            & = \frac{8 \sigma_H^4 N_r (N_t-1) + 4\sigma_W^2\sigma^2_H N_r}
                     {N_{rt}}
\end{align}
where we have used (\ref{Eq:Pre_Massive_MIMO_Eq23_2}),
(\ref{Eq:Pre_Massive_MIMO_Eq23_3}) and
(\ref{Eq:Pre_Massive_MIMO_Eq23_5}) and the fact that
\begin{equation}
\label{Eq:Pre_Massive_MIMO_Eq23_14}
 E
\left[
\tilde{U}_{k,\, i}'
\right] = 0
\end{equation}
where $\tilde{U}_{k,\, i}'$ is defined in (\ref{Eq:Pre_Massive_MIMO_Eq23_12}).

Noting that the average information content of $\tilde{Y}_i$ in
(\ref{Eq:Pre_Massive_MIMO_Eq23_11}) is $1/2$ bit, the average SINR per bit of
$\tilde{Y}_i$ is
\begin{align}
\label{Eq:Pre_Massive_MIMO_Eq23_15}
\mathrm{SINR}_{\mathrm{av},\, b,\, C}
& = \frac{E\left[\left|F_i S_i\right|^2\right]\times 2}
         {E\left[\left|\tilde{U}_i\right|^2\right]}             \nonumber  \\
& = \frac{8\sigma_H^2 (N_r N_{rt}+1)\times 2}
         {8\sigma_H^2  (N_t-1)+4\sigma^2_W}
\end{align}
where the subscript ``$C$'' denotes ``after combining'' and we have used
(\ref{Eq:Pre_Massive_MIMO_Eq8}) and (\ref{Eq:Pre_Massive_MIMO_Eq23_13}). When
$\sigma^2_W=0$ and $N_r N_{rt}\gg 1$, we get the upper bound as
\begin{align}
\label{Eq:Pre_Massive_MIMO_Eq23_16}
\mathrm{SINR}_{\mathrm{av},\, b,\, C,\,\mathrm{UB}}
& = \frac{8\sigma_H^2  (N_r N_{rt}+1)\times 2}
         {8\sigma_H^2  (N_t-1)}                          \nonumber  \\
& \approx
  \frac{2N_{rt}N_r}{N_t-1}                               \nonumber  \\
& \approx
  \mathrm{SINR}_{\mathrm{av},\, b,\,\mathrm{UB}}
\end{align}
for $N_r\gg 1$. Note that re-transmissions increases the upper bound on the
average SINR per bit, it does not improve the BER.
\section{Simulation Results}
\label{Sec:Results}
In this section, we discuss the results from computer simulations.
\begin{figure*}[tbhp]
\begin{center}
\input{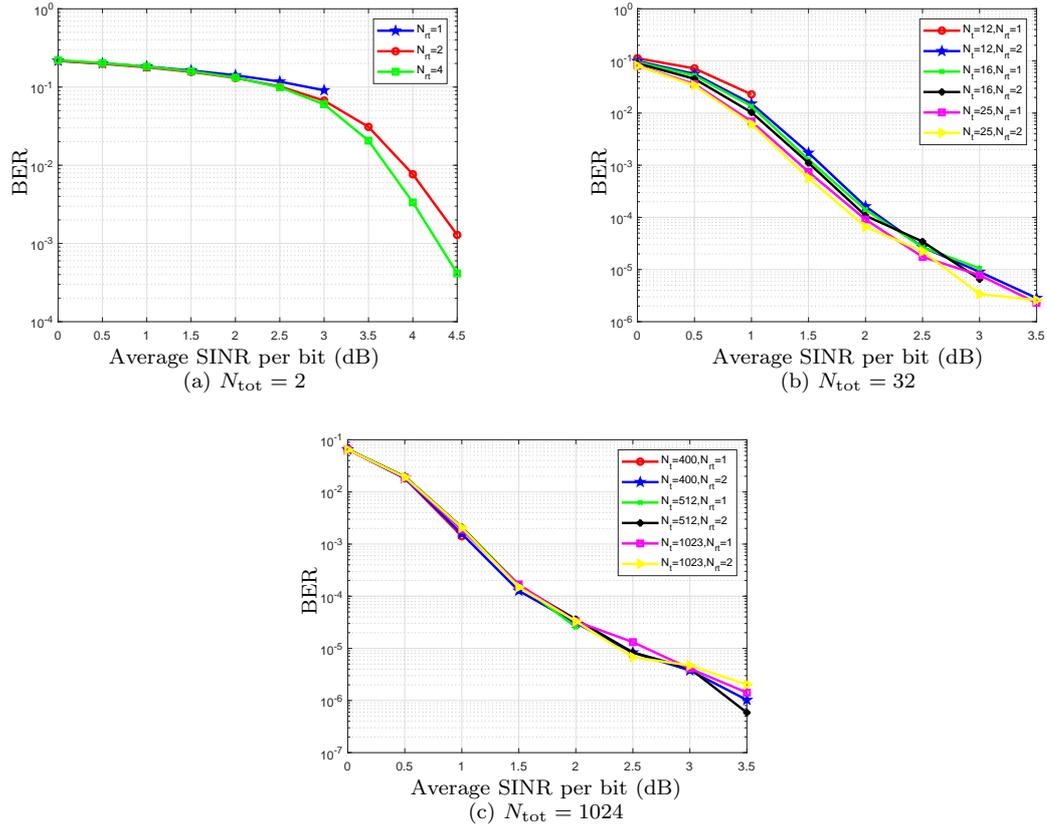}
\caption{Simulation results with precoding.}
\label{Fig:Pap14_Results}
\end{center}
\end{figure*}
\begin{figure*}[tbhp]
\begin{center}
\input{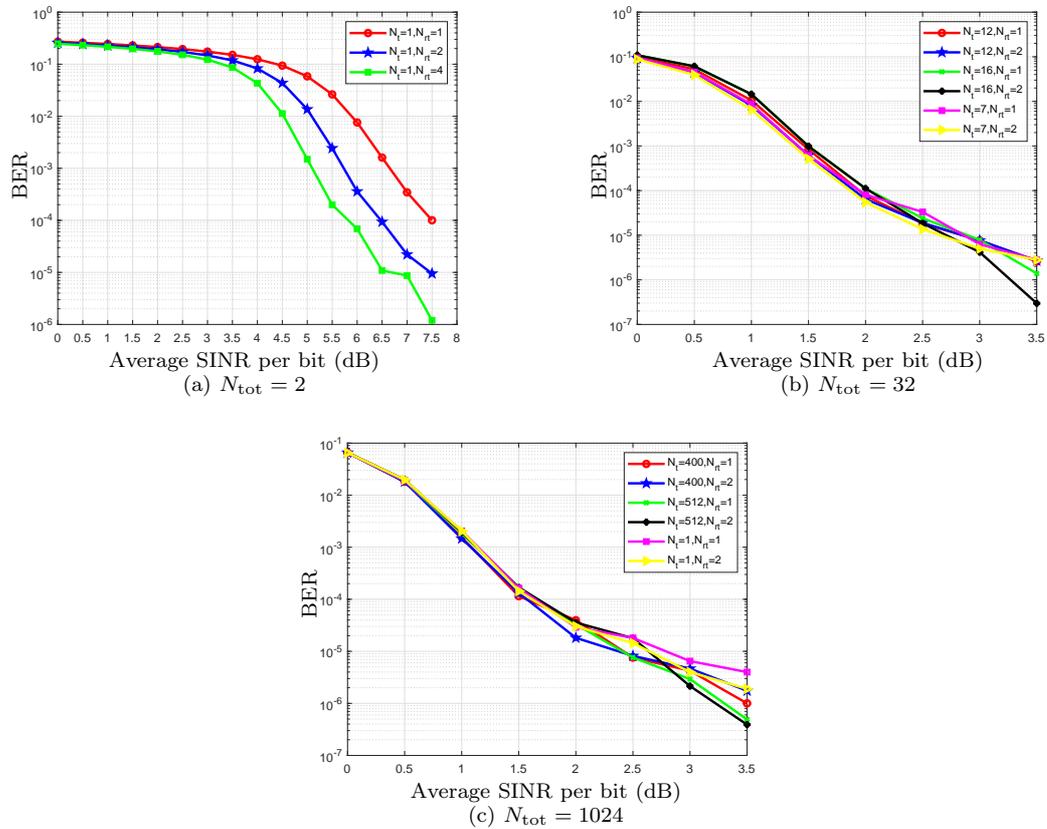}
\caption{Simulation results without precoding.}
\label{Fig:Pap16_Results}
\end{center}
\end{figure*}
For the precoded case, the length of the data bits per ``frame'' ($L_{d1}$)  is
taken to be the smallest integer greater than 1000, which is an integer multiple
of $N_r$. For the case without precoding, the length of the data bits per ``frame''
($L_{d1}$)  is taken to be the smallest integer greater than 1000, which is an
integer multiple of $N_t$. Note that (see \textbf{Figs.}~\ref{Fig:Pap14_System}
and \ref{Fig:No_Precode_System})
\begin{equation}
\label{Eq:Pre_Massive_MIMO_Eq24}
L_d = 2 L_{d1}.
\end{equation}
The simulations were carried out over $10^4$ frames. The turbo encoder is
given by (38) of \cite{KV_OpSigPJ2019}. Figure~\ref{Fig:Pap14_Results} gives the
bit-error-rate (BER) results with precoding, whereas Figure~\ref{Fig:Pap16_Results}
gives the BER performance without precoding.
\begin{itemize}
 \item \textbf{Fig.}~\ref{Fig:Pap14_Results}(a) gives the bit-error-rate (BER)
       results for a $1\times 1$ single input single output (SISO) system
       ($N_{\mathrm{tot}}=2$) with precoding. We get a BER of $2\times 10^{-2}$
       at an average
       SNR per bit of 3.5 dB, with $N_{rt}=4$. The corresponding spectral
       efficiency is $\eta^p=1/8$ bits per transmission. The BER also does not
       vary significantly with the number of re-transmissions ($N_{rt}$).
 \item \textbf{Fig.}~\ref{Fig:Pap14_Results}(b) gives the results for
       $N_{\mathrm{tot}}=32$ and different combinations of transmit ($N_t$)
       and receive ($N_r$) antennas. We find that the BER is quite insensitive
       to variations in $N_t$, $N_r$ and $N_{rt}$. Moreover, the BER at an
       SNR per bit of 3.5 dB is about $2\times 10^{-6}$, which is a significant
       improvement over the SISO system. Of all the curves, $N_t=25$,
       $N_{rt}=2$ gives the lowest spectral efficiency of $\eta^p=1.75$
       bits/sec/Hz and highest
       $\mathrm{SNR}_{\mathrm{av},\, b,\,\mathrm{UB}}=12.39$ dB. Of all the
       curves, $N_t=12$, $N_{rt}=1$ gives the highest spectral efficiency
       $\eta^p=10$ bits/sec/Hz and lowest
       $\mathrm{SNR}_{\mathrm{av},\, b,\, \mathrm{UB}}=1.36$ dB.
 \item \textbf{Fig.}~\ref{Fig:Pap14_Results}(c) gives the results for
       $N_{\mathrm{tot}}=1024$ for various combinations of $N_t$, $N_r$
       and $N_{rt}$. The BER is similar to that of $N_{\mathrm{tot}}=32$. Of all
       the curves, $N_t=400$, $N_{rt}=1$ gives the highest spectral efficiency
       of $\eta^p=312$ bits/sec/Hz and lowest
       $\mathrm{SNR}_{\mathrm{av},\, b,\,\mathrm{UB}}=1.09$ dB. Of all the
       curves, $N_t=1023$, $N_{rt}=2$ gives the lowest spectral efficiency of
       $\eta^p=0.25$ and highest
       $\mathrm{SNR}_{\mathrm{av},\, b,\,\mathrm{UB}}\rightarrow\infty$.
 \item \textbf{Fig.}~\ref{Fig:Pap16_Results}(a) gives the bit-error-rate (BER)
       results for a $1\times 1$ single input single output (SISO) system
       ($N_{\mathrm{tot}}=2$) without precoding. We get a BER of $10^{-1}$ at
       an average SNR per bit of 3.5 dB, with $N_{rt}=4$. The corresponding spectral
       efficiency is $\eta=1/8$ bits per transmission. Compared to \textbf{Fig.}~3(a)
       of \cite{KV_OpSigPJ2019}, we find that the BER varies significantly with the
       number of re-transmissions ($N_{rt}$) in \textbf{Fig.}~\ref{Fig:Pap16_Results}(a)
       of this work. However, it must be noted that the definition of
       $\mathrm{SNR}_{\mathrm{av},\, b}$ in (23) of \cite{KV_OpSigPJ2019} and
       $\mathrm{SINR}_{\mathrm{av},\, b}$ (converted to decibels) in
       (\ref{Eq:Pre_Massive_MIMO_Eq23_6}) of this work, are different. Comparing
       (23) of \cite{KV_OpSigPJ2019} and (\ref{Eq:Pre_Massive_MIMO_Eq23_6}) (converted
       to decibels) in this work for $N_t=N_r=1$, we find that
       (\ref{Eq:Pre_Massive_MIMO_Eq23_6})
       is 3 dB higher. This explains the 3 dB difference between \textbf{Fig.}~3(a)
       of \cite{KV_OpSigPJ2019} and \textbf{Fig.}~\ref{Fig:Pap16_Results}(a) of
       this work.
 \item \textbf{Fig.}~\ref{Fig:Pap16_Results}(b) gives the results for
       $N_{\mathrm{tot}}=32$ and different combinations of transmit ($N_t$)
       and receive ($N_r$) antennas. We find that the BER is quite insensitive
       to variations in $N_t$, $N_r$ and $N_{rt}$. Moreover, the BER at an
       SNR per bit of 3.5 dB is about $10^{-6}$, which is a significant
       improvement over the SISO system and similar to the precoded system of
       \textbf{Fig.}~\ref{Fig:Pap14_Results}(b). Of all the curves, $N_t=7$,
       $N_{rt}=2$ gives the lowest spectral efficiency of $\eta=1.75$
       bits/sec/Hz and highest
       $\mathrm{SNR}_{\mathrm{av},\, b,\,\mathrm{UB}}=12.39$ dB. Of all the
       curves, $N_t=16$, $N_{rt}=1$ gives the highest spectral efficiency equal to
       $\eta=8$ bits/sec/Hz and lowest
       $\mathrm{SNR}_{\mathrm{av},\, b,\, \mathrm{UB}}=5.4$ dB.
 \item \textbf{Fig.}~\ref{Fig:Pap16_Results}(c) gives the results for
       $N_{\mathrm{tot}}=1024$ for various combinations of $N_t$, $N_r$
       and $N_{rt}$. The BER is similar to that of $N_{\mathrm{tot}}=32$. Of all
       the curves, $N_t=512$, $N_{rt}=1$ gives the highest spectral efficiency
       of $\eta=256$ bits/sec/Hz and lowest
       $\mathrm{SNR}_{\mathrm{av},\, b,\,\mathrm{UB}}=3.03$ dB. Of all the
       curves, $N_t=1$, $N_{rt}=2$ gives the lowest spectral efficiency of
       $\eta=0.25$ bits/sec/Hz and highest
       $\mathrm{SNR}_{\mathrm{av},\, b,\,\mathrm{UB}}\rightarrow\infty$.
\end{itemize}
\section{Conclusions}
\label{Sec:Conclude}
This work presents an elegant method for data detection in turbo-coded
massive MIMO with and without precoding. An ideal receiver is assumed. Simulation
results show that the BER is quite insensitive to a wide range in the number of
transmit antennas and re-transmissions, when the total number of antennas in the
transmitter and receiver ($N_{\mathrm{tot}}$) is large and kept constant. Thus,
the spectral efficiency
can be made as large as possible for a given BER and $N_{\mathrm{tot}}$. Future work
could be to simulate a realistic massive MIMO system with carrier and timing
synchronization and channel estimation.
\begin{footnotesize}

\bibliographystyle{IEEEtran}
\bibliography{mybib,mybib1,mybib2,mybib3,mybib4,mybib5}
\end{footnotesize}
\end{document}